\documentclass[10pt,onecolumn]{IEEEtran}

\usepackage{graphicx}
\usepackage{float}
\usepackage{times}
\usepackage{amsmath}
\usepackage{amsfonts}
\usepackage{balance}
\usepackage{multirow}

\renewcommand{\baselinestretch}{1.1}

\begin{document}
\title{Compressive Sensing Based Design of Sparse Tripole Arrays}


\author{\IEEEauthorblockN{Matthew Hawes $^{1,}$, Wei Liu $^{2}$ and Lyudmila Mihaylova $^{1}$}\\[0.3cm]
\IEEEauthorblockA{$^{1}$ Department of
Automatic Control and Systems Engineering, University of Sheffield, S1 3JD, UK\\$^{2}$ Department of Electronic and Electrical Engineering, University of Sheffield, S1 3JD, UK\\
{\{m.hawes, l.s.mihaylova, w.liu\}@sheffield.ac.uk }}

}


\maketitle

\begin{abstract}
This paper considers the problem of designing sparse linear tripole arrays.  In such arrays at each antenna location there are three orthogonal dipoles, allowing full measurement of both the horizontal and vertical components of the received waveform.  We formulate this problem from the viewpoint of Compressive Sensing (CS).  However, unlike for isotropic array elements (single antenna), we now have three complex valued weight coefficients associated with each potential location (due to the three dipoles), which have to be simultaneously minimised.  If this is not done, we may only set the weight coefficients of individual dipoles to be zero valued, rather than complete tripoles, meaning some dipoles may remain at each location.  Therefore, the contributions of this paper are to formulate the design of sparse tripole arrays as an optimisation problem, and then we obtain a solution based on the minimisation of a modified $l_{1}$ norm or a series of iteratively solved reweighted minimisations, which ensure a truly sparse solution.  Design examples are provided to verify the effectiveness of the proposed methods and show that a good approximation of a reference pattern can be achieved using fewer tripoles than a Uniform Linear Array (ULA) of equivalent length.
\end{abstract}
%%%%%%%%%%%%%%%%%%%%%%%%%%%%%%%%%%%%%%%%%%%%%%%
\section{Introduction}

Sensor arrays are used in a wide range of application areas including Direction Of Arrival (DOA) estimation and beamforming \cite{vantrees02a}, vehicle health maintenance \cite{Rodger12} and many others.  When considering fixed beamformer design, it is well known that an adjacent antenna separation of less than half a wavelength is required if using Uniform Linear Arrays (ULAs) in order to avoid the introduction of grating lobes \cite{vantrees02a}.  This can become problematic when considering arrays with large apertures due to the cost associated with the number of antennas required.  As a result, sparse arrays have become a desirable alternative as they allow a given aperture to be implemented using fewer antennas.  The reason for this is that their nonuniform adjacent antenna separations allow mean separations greater than half of the wavelength, while still avoiding the introduction of grating lobes \cite{Jarske88}.

The tradeoff is that the unpredictable sidelobe behaviour of sparse arrays means that some optimisation of antenna locations is required in order to ensure an acceptable performance is achieved.  For example, we could consider minimising the Peak Sidelobe Level (PSL) or matching the response to an acceptable reference response.  Stochastic optimisation methods such as Genetic Algorithms (GAs) \cite{Haupt94,Yan97a,Chen06B,Cen08A} and Simulated Annealing (SA) \cite{Trucco99a,Repetto06} have been shown to be able to solve such a problem.  Improved performances can also be achieved by combining GAs with Difference \mbox{Sets (DS) \cite{Caorsi04}} and Almost Difference Sets (ADS) \cite{Oliveri11}.  However, the disadvantages of GAs and similar methods are the potentially long computation times and the possibility of failing to converge to the optimal solution.

Instead, Compressive Sensing (CS) \cite{Candes06} (or alternatively the Bayesian formulation \cite{Ji08}) tells us that, when certain conditions are met, we can recover signals from fewer measurements than are traditionally required.  CS-based methods have been applied to various array signal processing problems such as DOA estimation \cite{Wang13} and in the design of sparse antenna arrays \cite{Li08,Carin09,Cen10b,Prisco11,Oliveri12}.  For CS-based sparse array design methods, the aim is to use as few antennas as possible, while still getting an exact, or almost exact, match to a reference response.  It has further been shown that the sparsity of the solution can be improved by converting the problem into a series of iteratively solved reweighted minimisations \cite{Fuchs12,Prisco12}.  This is achieved by making the problem a closer approximation of an $l_{0}$ norm minimisation by adding a reweighting term that penalises small weight coefficients more heavily \cite{Candes08}.

Previous work involving CS and sparse array design has tended to focus on the case of isotropic array elements.  This means that the polarisation of the signal of interest is not accounted for, potentially degrading the performance of an array.  Alternatively, we can consider arrays consisting of vector sensors allowing measurement of different components of the received electromagnetic waveform \cite{WilliamsDSP,Zhong14,Zheng15,Wang15}.  One such vector sensor is a tripole, which consists of three orthogonally orientated dipoles \cite{Compton81b,Lep98,Bansal,Wong01a,Wong01b,Lund04,A-Y09,Yuan12,Getu13a}.  Such vector sensors have been used in various configurations for DOA and polarisation estimation \cite{Zolt00a,Zolt00b,Wong00a,Wong04,Wong11a,He12,Zheng14,Song14a}.  However, in these works, fixed sensor locations are considered rather than addressing the location optimisation problem associated with sparse array design.

The novelty of this paper arises in the proposed solution to the problem of designing sparse tripole arrays using CS for the first time.  Unlike for isotropic array elements (consisting of array elements of a single antenna), we now have three complex valued weight coefficients (one for each co-located dipole) that have to be simultaneously minimised for each potential tripole location.  If this is not done, there is the potential for only the coefficients of single dipoles (rather than complete tripoles) to be made zero-valued.  As a result, we can obtain a solution with at least one dipole at each location on the grid of potential sensor locations rather than obtaining a truly sparse solution.  Instead, to ensure the desired sparsity is achieved, we propose two CS-based methods of solving this problem: (i) the optimisation can be formulated as a modified $l_{1}$ norm minimisation; (ii) the sparsity can be further improved by converting the problem into a series of iteratively solved reweighted minimisations.  We provide broadside and off-broadside design examples to show that the proposed methods can successfully obtain a good approximation of a reference pattern using fewer tripoles than a ULA of equivalent length.

This work differs from previous work of the first and third authors as follows: (i) In \cite{Hawes12,Hawes14b,Hawes14c}, we considered the problem of designing sparse sensor arrays for both narrowband and wideband beamforming.  Unlike the current work, the arrays consisted of isotropic array elements with a single sensor at each location.  In \cite{Hawes15b}, we dealt with the design of sparse linear arrays based on crossed dipoles.  However, \cite{Hawes15b} deals with a quaternionic signal model, which is not utilised in this work; \mbox{(ii) \cite{Zhang13} deals with} DOA estimation using a fixed spatially stretched tripole (no co-located tripole) rather than the sparse array design problem.

The remainder of this paper is structured as follows,
Section \ref{sec:design} gives details of the proposed design methods including: the array model in Section \ref{sub:AM}  and the CS based formulations in Section \ref{sub:CS}. Design examples are presented in Section \ref{sec:sim}, with conclusions being drawn in Section \ref{sec:con}.

%%%%%%%%%%%%%%%%%%%%%%%%%%%%%%%%%%%%%%%%%%
\section{Proposed Design Methods}\label{sec:design}
\unskip
\subsection{Array Model}\label{sub:AM}

\begin{figure}[H]
\begin{center}
   \includegraphics[angle=0,width=.7\textwidth]{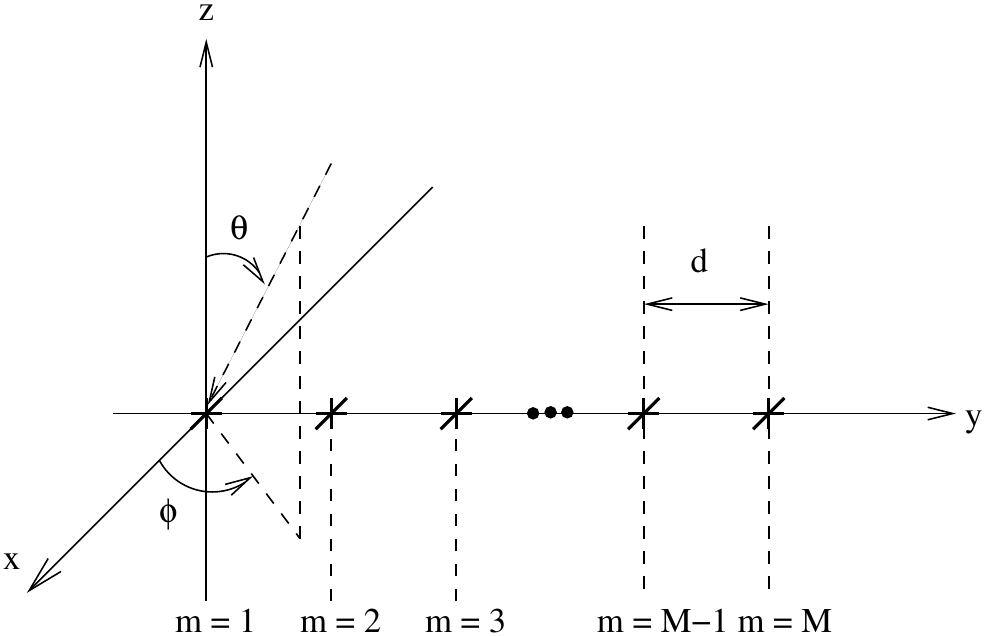}
   \caption{An array model with $M$ potential tripole locations shown.
    \label{fig:AM}}
\end{center}
\end{figure}

Figure \ref{fig:AM} shows the tripole array structure being
considered.  There are $M$ tripole locations spread over the y-axis with an adjacent separation distance $d$.  For each tripole, there are three orthogonally orientated dipoles, one parallel to each axis.  Also shown is a signal with its DOA defined by the angles $\theta$ and $\phi$, which are limited as follows: $0\leq\theta\leq\pi/2$ and $-\pi/2\leq\phi\leq\pi/2$.  A
plane-wave signal model is assumed, \textit{i.e.}, the signal impinges upon
the array from the far field.

The spatial steering vector of the array is given by
\begin{eqnarray}\nonumber
\label{eq:s_s}
    \textbf{s}_{s}(\theta,\phi)&=&[1, e^{-j2\pi d\sin\theta\sin\phi/\lambda},\\ &&\ldots, e^{-j2\pi(M-1)d\sin\theta\sin\phi/\lambda}]
\end{eqnarray}
where $\lambda$ is the wavelength of the signal of interest.  For tripoles,
the spatial-polarization coherent vector contains information about
a signal's polarisation and is given by
\cite{Compton81b}:
\begin{eqnarray}\label{eq:s_p}\nonumber
\textbf{s}_{p}(\theta,\phi,\gamma,\eta) &=&
\left[
  \begin{array}{c}
    \sin\gamma\cos\theta\cos\phi e^{j\eta} - \cos\gamma\sin\phi \\
    \sin\gamma\cos\theta\sin\phi e^{j\eta} - \cos\gamma\cos\phi \\
    -\sin\gamma\sin\theta e^{j\eta} \\
  \end{array}
\right] \\  &=& \left[
  \begin{array}{c}
    s_{p,x}(\theta,\phi,\gamma,\eta) \\
    s_{p,y}(\theta,\phi,\gamma,\eta) \\
    s_{p,z}(\theta,\phi,\gamma,\eta) \\
  \end{array}
\right]
 \end{eqnarray}
where $0\leq\gamma\leq\pi/2$ is the auxiliary polarization angle and
$-\pi\leq\eta<\pi$ is the polarization \mbox{phase difference.}

Now, the array can be split into three sub-arrays, one parallel to each axis.  The steering vector of each of these sub-arrays is complex-valued and given by:
\begin{equation}\label{eq:s_x}
\textbf{s}_{x}(\theta,\phi,\gamma,\eta) =
 s_{p,x}(\theta,\phi,\gamma,\eta)\textbf{s}_{s}(\theta,\phi)
 \end{equation}
\begin{equation}\label{eq:s_y}
\textbf{s}_{y}(\theta,\phi,\gamma,\eta) =
 s_{p,y}(\theta,\phi,\gamma,\eta)\textbf{s}_{s}(\theta,\phi)
 \end{equation}
 and
\begin{equation}\label{eq:s_z}
\textbf{s}_{z}(\theta,\phi,\gamma,\eta) =
 s_{p,z}(\theta,\phi,\gamma,\eta)\textbf{s}_{s}(\theta,\phi)
 \end{equation}
 respectively.

 The response of the array is given by
\begin{equation}\label{eq:p2}
  p(\theta,\phi,\gamma,\eta)=\textbf{s}(\theta,\phi,\gamma,\eta)\textbf{w}^{H}
\end{equation}
with
\begin{equation}\label{eq:w1}
  \textbf{w} = [w_{x,1},w_{y,1},w_{z,1}, \ldots, w_{x,M},w_{y,M},w_{z,M}]
\end{equation}
where $w_{i,m}$ is the weight coefficient for the $m^{th}$ dipole orientated parallel to the $i$-axis ($i=\{x,y,z\}$) and $\{.\}^{H}$ denotes the Hermitian transpose.  Similarly, we also have
\begin{eqnarray}\label{eq:s1}\nonumber
  \textbf{s}(\theta,\phi,\gamma,\eta) &=& [s_{x,1}(\theta,\phi,\gamma,\eta),s_{y,1}(\theta,\phi,\gamma,\eta),\\ \nonumber &&s_{z,1}(\theta,\phi,\gamma,\eta), \ldots, s_{x,M}(\theta,\phi,\gamma,\eta),\\ &&s_{y,M}(\theta,\phi,\gamma,\eta),s_{z,M}(\theta,\phi,\gamma,\eta)]
\end{eqnarray}
where $s_{i,m}(\theta,\phi,\gamma,\eta)$ is the contribution of the $m^{th}$ dipole to the overall steering vector parallel to \mbox{the $i$-axis.}

In what follows, we address the problem of designing a sparse linear tripole array, where there are two issues to be solved.  Firstly, we have to find the optimal tripole locations, as there will no longer be a uniform adjacent tripole separation of $d$.  Secondly, we have to find the weight coefficients that give an acceptable beam response.  We propose to use design methods based on CS to solve \mbox{these problems.}

\subsection{Compressive Sensing Based Design of Sparse Tripole Arrays}\label{sub:CS}

Suppose $P_{r}(\theta,\phi,\gamma,\eta)$ is a reference response
which we wish to achieve. First, consider Figure \ref{fig:AM} as being
a grid of potential tripole locations.  In this
instance, $(M-1)d$ is the potential aperture of the array and $M$ is a large
number. Sparseness is then introduced by selecting the weight
coefficients to give as few active tripoles as possible, or, in other words, as few non-zero valued weight coefficients as possible.  This has to be done
while still giving a designed response that is close to the desired
one.

This problem is formulated as
\begin{eqnarray}\label{eq:min1}\nonumber
    &\min&||\textbf{w}||_{1}\\ &\textrm{subject to}&
    ||\textbf{p}_r-\textbf{S}\textbf{w}^{H}||_{2}\leq\alpha\;
\end{eqnarray}
where $||\textbf{w}||_{1}$ is the $l_{1}$ norm of the weight coefficients and used as an estimate of the number of nonzero weight coefficients in $\textbf{w}$, \textit{i.e.}, an approximation to the $l_{0}$ norm, $\textbf{p}_r$ is the vector holding
the desired beam response at the sampled angular and polarisation
points of interest, $\textbf{S}$ is the matrix composed of the
corresponding steering vectors and $\alpha$ places a limit on the
allowed difference between the desired and the designed responses.  Therefore, by minimising $||\textbf{w}||_{1}$ in Equation \eqref{eq:min1}, we are minimising the number of tripoles we have to use, while the constraint ensures that we still get a good approximation of the reference response.

In detail, $\textbf{p}_r$ and $\textbf{S}$ are respectively
given by
\begin{eqnarray}
\label{eq:pr}
    \textbf{p}_r&=&[P_r(\theta_{1},\phi_{1},\gamma_{1},\eta_{1}), \ldots, P_r(\theta_{L},\phi_{L},\gamma_{L},\eta_{L})]^{T}\\
\label{eq:S}
    \textbf{S}&=&[\textbf{s}(\theta_{1},\phi_{1},\gamma_{1},\eta_{1}),
    \ldots,
    \textbf{s}(\theta_{L},\phi_{L},\gamma_{L},\eta_{L})]^{T}
\end{eqnarray}
where $L$ is the number of points sampled at each dimension of the
desired beam response and
$\{.\}^{T}$ denotes the transpose operation.  In this work, $\textbf{p}_{r}$ is the ideal
response, \textit{i.e.}, a value of one for the mainlobe and zeros for the
other entries.  It would also be possible to use other reference responses if desired, for example, a response which defines a maximum sidelobe level or one with nulls at a \mbox{given direction.}

Note, we have to choose $L$ to be large enough to ensure that all angular and polarisation points of interest are being considered.  However, the larger $L$ is, the longer the computation time will be.  An acceptable compromise seems to be to sample the sidelobe regions every $1^{\circ}$.  We can also help improve the sparsity of the solution by increasing the value of $M$.  A larger value of $M$ gives a denser grid which is more likely to include the optimal locations.  As a result, the same amount of error between designed and reference responses can be obtained using fewer tripoles.  However, we will get to the point where the optimal locations will be already included on the grid and further increases will only serve to lengthen the computation time.  This means there is again a compromise to be achieved when selecting the value of $M$, and this will be explored in the design examples we present \mbox{in this paper.}

One problem with the above formulation is that the three complex weight coefficients associated with each tripole can not be guaranteed to be minimised simultaneously.  As a result, we may not be able to obtain a sparse solution even if we have a sparse weight vector $\textbf{w}$ after the minimisation.  In other words, it would be possible to end up only removing a single dipole from each potential location rather than truly introducing sparsity by removing complete tripoles.  Therefore, we further modify the formulation in Equation (\ref{eq:min1}) by converting it into a modified $l_{1}$ norm minimisation.

First, we rewrite Equation (\ref{eq:min1}) as
\begin{eqnarray}\label{eq:mint}\nonumber
    &\min& t \;\; \epsilon\;\;  {R}^{+}\\ &\textrm{subject to}&
    ||\textbf{p}_{r}-\textbf{S}\textbf{w}^{H}||_{2}\leq\alpha\;, \;\;\;\;|\langle\textbf{w}\rangle|_{1}\leq t
\end{eqnarray}
where
\begin{equation}\label{eq:constraint2}
    |\langle\textbf{w}\rangle|_{1}=\sum_{m=1}^{M}||\textbf{w}_{m}||_{2}
\end{equation}
and $\textbf{w}_{m} = [w_{x,m},w_{y,m},w_{z,m}]$.  Here $\textbf{w}_{m}$ contains the weight coefficients for all the dipoles that make up the tripole at the $m^{th}$ grid location.  Therefore, by minimising the value of $t$ the second constraint in Equation \eqref{eq:mint} ensures the combined weight coefficients for the tripoles are also minimised.

Now, we decompose $t$ to $t=\sum_{m=1}^{M}t_{m}$,
$t_{m}\epsilon\;\; {R}^{+}$.  This allows us to minimise the combined weight coefficients for each tripole separately. In vector form, we have
\begin{equation}
  t=[1, \cdots, 1]\left[
                                                     \begin{array}{c}
                                                       t_{1} \\
                                                       \vdots \\
                                                       t_{M} \\
                                                     \end{array}
                                                   \right]=\textbf{1}^T\textbf{t}.
\end{equation}
Then, Equation (\ref{eq:mint}) can be rewritten as
\begin{eqnarray}\label{eq:min1t}\nonumber
&\min\limits_{\textbf{t}}& \textbf{1}^T\textbf{t}\\ \nonumber
&\textrm{subject to}&
    ||\textbf{p}_{r}-\textbf{S}\textbf{w}^{H}||_{2}\leq\alpha\\ &&
    ||\textbf{w}_{m}||_{2}\leq t_{m},\; \;m=1, \cdots,M.
\end{eqnarray}

Now, define
\begin{eqnarray}\label{eq:w_hat}\nonumber
  \hat{\textbf{w}} &=& [t_{1}, R(w_{x,1}), R(w_{y,1}), R(w_{z,1}), -I(w_{x,1}), -I(w_{y,1}),\\ && -I(w_{z,1}), t_{2}, \ldots, -I(w_{z,M})],\\
\label{eq:c_hat}
    \hat{\textbf{c}} &=& [1, 0, 0, 0, 0, 0, 0, 1, 0, 0, \cdots, 1, 0, 0, 0, 0, 0, 0]^{T}\;,\\
    \label{eq:prhat}
    \hat{\textbf{p}}_{r} &=& [\textbf{R}(\textbf{p}_{r}),\textbf{I}(\textbf{p}_{r})]^{T}
\end{eqnarray}
and
\begin{equation}\label{eq:S_hat}
    \hat{\textbf{S}}=\left(
                       \begin{array}{cc}
                         \boldsymbol{0} & \boldsymbol{0}  \\
                         \textbf{R}(\textbf{s}_{x,1}) & \textbf{I}(\textbf{s}_{x,1})  \\
                         \textbf{R}(\textbf{s}_{y,1}) & \textbf{I}(\textbf{s}_{y,1})  \\
                         \textbf{R}(\textbf{s}_{z,1}) & \textbf{I}(\textbf{s}_{z,1})  \\
                         -\textbf{I}(\textbf{s}_{x,1}) & \textbf{R}(\textbf{s}_{x,1})  \\
                         -\textbf{I}(\textbf{s}_{y,1}) & \textbf{R}(\textbf{s}_{y,1})  \\
                         -\textbf{I}(\textbf{s}_{z,1}) & \textbf{R}(\textbf{s}_{z,1})  \\
                         \vdots & \vdots  \\
                         \boldsymbol{0} & \boldsymbol{0} \\
                         \vdots & \vdots  \\
                         -\textbf{I}(\textbf{s}_{z,M}) & \textbf{R}(\textbf{s}_{z,M})  \\
                       \end{array}
                     \right)^{T}
\end{equation}
where $\textbf{s}_{i,m}$ is the contribution of the $m^{th}$ dipole parallel to the $i-$axis to the steering vector of the array, $R(.)$ gives the real component of a scalar and $I(.)$ gives the imaginary component of a scalar and $\textbf{R}(.)$ gives the real components of a vector/matrix and $\textbf{I}(.)$ gives the imaginary components of a vector/matrix.  Here, we have included the values of $t_{m}$ in $\hat{\textbf{w}}$ as we want to automatically find them with the weight coefficients instead of them being predetermined.  We then use $\hat{\textbf{c}}$ to select the values of $t_{m}$ for minimisation, and the zeros are added in $\hat{\textbf{S}}$ to ensure they do not contribute to the error between the reference and designed responses.

Then, the final formulation is as follows
\begin{eqnarray}\label{eq:cw}\nonumber
&\min\limits_{\hat{\textbf{w}}}& \hat{\textbf{w}}\hat{\textbf{c}}\\
\nonumber &\textrm{subject to}&
    ||\hat{\textbf{p}}_{r}-\hat{\textbf{S}}\hat{\textbf{w}}^{T}||_{2}\leq\alpha\\&&
    ||\textbf{w}_{m}||_{2}\leq t_{m},\; \;m=1, \cdots, M
\end{eqnarray}
which can be solved using cvx, a package for specifying and solving convex programs \cite{cvx,Grant08}.

As with the traditional CS based design methods, the solution can be improved by considering the problem as a series of iteratively solved reweighted minimisations to bring the solution closer to that of the $l_{0}$ norm minimisation \cite{Candes08,Fuchs12,Prisco12}.  This is achieved by the addition of a reweighting term which is found from the previous iteration and penalises small non-zero valued weight coefficients more heavily, meaning all non-zero valued coefficients are treated in a more uniform manner (as for the $l_{0}$ norm minimisation).  As a result, these small non-zero valued weight coefficients are less likely to be repeated in the next iteration and the sparsity of the solution is improved.

In this form, the problem is formulated as follows
\begin{eqnarray}\label{eq:cw2}\nonumber
&\min\limits_{\hat{\textbf{w}}}& \hat{\textbf{w}}\hat{\textbf{c}}\\
\nonumber &\textrm{subject to}&
    ||\hat{\textbf{p}}_{r}-\hat{\textbf{S}}\hat{\textbf{w}}^{T}||_{2}\leq\alpha\\&&
    \delta_{m}^{k}||\textbf{w}_{m}||_{2}\leq t_{m},\; \;m=1, \cdots, M
\end{eqnarray}
where we now have
\begin{equation}\label{eq:c_hatre}
    \hat{\textbf{c}} = [\delta^{k}_{1}, 0, 0, 0, 0, 0, 0, \delta^{k}_{2}, 0, \ldots, \delta^{k}_{M},0, 0, 0, 0, 0, 0]^{T}
\end{equation}
and
\begin{equation}\label{eq:a}
    \delta^{k}_{m} = (|\textbf{w}_{m}^{k-1}|+\epsilon)^{-1}.
\end{equation}
This is then solved iteratively until the number of non-zero valued weight coefficients has remained constant for a few iterations.

Note, the term $\epsilon$ is added for numerical stability and should be set slightly smaller than the minimum implemented weight coefficient for a single tripole.  However, its addition means that a zero-valued weight coefficient in one iteration will not be guaranteed to be repeated in the next.  This has the result of meaning a solution is not always guaranteed, but when a solution is possible, it will usually be achieved in less than 10 iterations.  For the first iteration, Equation (\ref{eq:cw}) can be solved and the reweighting terms introduced from the second iteration onwards.  Finally, $\textbf{w}_{m}$ gives the current estimate of the weight coefficients for the $m^{th}$ tripole, whereas $\textbf{w}_{m}^{k-1}$ gives the estimate from the previous iteration.

%%%%%%%%%%%%%%%%%%%%%%%%%%%%%%%%%%%%%%%%%%

\section{Design Examples}\label{sec:sim}

We will now consider broadside and off-broadside design examples to verify the effectiveness of the two design methods.  All examples are implemented on a computer with
an Intel Xeon CPU E3-1271 (3.60 GHz), 16GB of RAM and Windows 7 operating system, using Matlab R2014a and \mbox{cvx \cite{cvx,Grant08}.}  In all of the figures that follow positive values of $\theta$ indicate the value range $\theta\in[0^\circ,\; 90^\circ]$ with
$\phi=90^{\circ}$, while negative values of $\theta\in[-90^\circ,\;
0^\circ]$ indicate an equivalent range of $\theta\in[0^\circ,\;
90^\circ]$ with $\phi=-90^{\circ}$.

\subsection{Broadside Design Examples}

First, we will consider broadside ($\theta_{ML}=0^{\circ},\phi_{ML}=90^{\circ}$) design examples for both methods and show how the value of $M$ effects their performances.  The length of the array is set to be $10\lambda$ and the values $\alpha=0.5, \gamma=55^{\circ}$ and $\eta=100^{\circ}$ selected.  Finally, the sidelobe regions are defined as $\theta_{SL}=[10^{\circ},90^{\circ}]$ for $\phi_{SL}=\pm90^{\circ}$.

Table \ref{tb:broadcompare} compares the performances of the proposed methods for values of $M=101, 301$ and $501$.  For each case, the results for the non-reweighted example are found using Equation \eqref{eq:cw} and for the reweighted example Equation \eqref{eq:cw2} is used.  Here, we are comparing the aperture of the designed array, number of tripoles required, mean adjacent tripole separation ($\overline{\Delta\textrm{$d$}}$), the minimum adjacent tripole separation, the computation time, designed mainlobe location ($\hat{\theta}_{ML}$) and PSL.  The first three give an indication of the level of sparsity introduced by the design methods, the computation time is used to indicate the implementation complexity and the final two measures give an indication of how desirable the final response is.  Note, we are not using $||\textbf{p}_r-\textbf{S}\textbf{w}^{H}||_{2}$ as a performance measure as the same value of $\alpha$ is used in each case, meaning the value of $||\textbf{p}_r-\textbf{S}\textbf{w}^{H}||_{2}$ will be the same for both methods.

\begin{table}[H]
\caption{\rm Broadside performance comparison.} \centering
\footnotesize
\begin{tabular}{cccccccc}
  &  & \textbf{Number of} &  & &\textbf{Computation}   & &\\
\textbf{Example}&\boldmath\textbf{Aperture}$/\lambda$ & \boldmath\textbf{Tripoles}& \boldmath$\overline{\Delta\textrm{$d$}}/\lambda$& \boldmath\textbf{Min(}$\Delta\textrm{$d$}$\textbf{)}& \textbf{Time (Seconds)}&\boldmath$\hat{\theta}_{ML}$ & \textbf{PSL (dB)}\\
\hline
 Non-reweighted & &  & & & &&\\

M= 101 & 10 & 16 & 0.67&0.15 & 5.61     &$90^{\circ}$ &$-$32.28\\
\hline
Reweighted  & &  & & & & &\\

M= 101  &  5.20& 8 &0.74 &0.60& 25.92  & $90^{\circ}$&$-$23.25\\
\hline
 Non-reweighted & &  & & && &\\

M= 301 &10  &  14& 0.77&0.57 &  23.86    & $90^{\circ}$&$-$32.53\\
\hline
Reweighted  & &  & & & & &\\

M= 301  & 5.13 & 8 & 0.73&0.60 &88.88  & $90^{\circ}$&$-$22.94\\
\hline

 Non-reweighted & &  & && & &\\

M= 501 & 10 & 14 & 0.77&0.58  &49.90    & $90^{\circ}$& $-$32.69\\
\hline
Reweighted  & &  & & & && \\

M= 501  & 5.12 & 8 &0.73 &0.60 & 168.52 & $90^{\circ}$&$-$22.89 \\

\end{tabular}
\label{tb:broadcompare}
\end{table}

The first thing to note is that, for all values of $M$, both methods have successfully obtained a good approximation of the reference pattern and achieved a desirable beam response.  This is evident from the fact that the mainlobes are always in the correct location with sufficient sidelobe attenuation simultaneously achieved.

When looking at the results for the non-reweighted examples, we can see that it is possible, but not guaranteed, that increasing the value of $M$ can help increase the sparsity.  This can be explained by considering that the denser the grid of potential tripole locations, the more likely it is that the grid will include the optimal tripole locations.  As a result, it would be possible to achieve the same error between reference and designed responses using fewer tripoles.  However, we will get to the point where the grid is dense enough to include the optimal locations and further increasing the value of $M$ has little or no effect.  This is evident from the observation that increasing $M$ from $101$ to $301$ has increased the mean and minimum adjacent separations as well as decreasing the number of tripoles required.  In other words, the sparsity has been increased.  However, when increasing $M$ further to $501$ only the minimum adjacent sensor separation has increased, showing less of an improvement has been achieved.  It is also worth noting that increasing $M$ has been shown to always increase the computation time.  If we take the computation time for $M=101$ as a reference, we can see that there is a $425\%$ increase when $M=301$ and a $978\%$ increase in computation time when $M=501$.  Similarly, for the reweighted examples, we have increases of $343\%$ for $M=301$ and $650\%$ for $M=501$.   As a result, a balance has to be achieved rather than continually increasing $M$.  In this case, $M=301$ seems to be a good choice.

From the results for $M=101$, we can see that using the reweighted method has improved all measures of sparsity of the result (\textit{i.e.}, number of tripoles, mean and minimum adjacent tripole separation).  For $M=301$ and $M=501$ we can see that the reweighted method still gives a significant improvement in the number of tripoles that are required.  As a result, a greater cost saving is made.  However, the iterative nature of this method means that in each case there has been an increase in the computation time, as would be expected.  For $M=101$ we have a $462\%$ increase, $373\%$ for $M=301$ and $338\%$ for $M=501$.

\begin{figure}[H]
\begin{center}
   \includegraphics[angle=0,width=.6\textwidth]{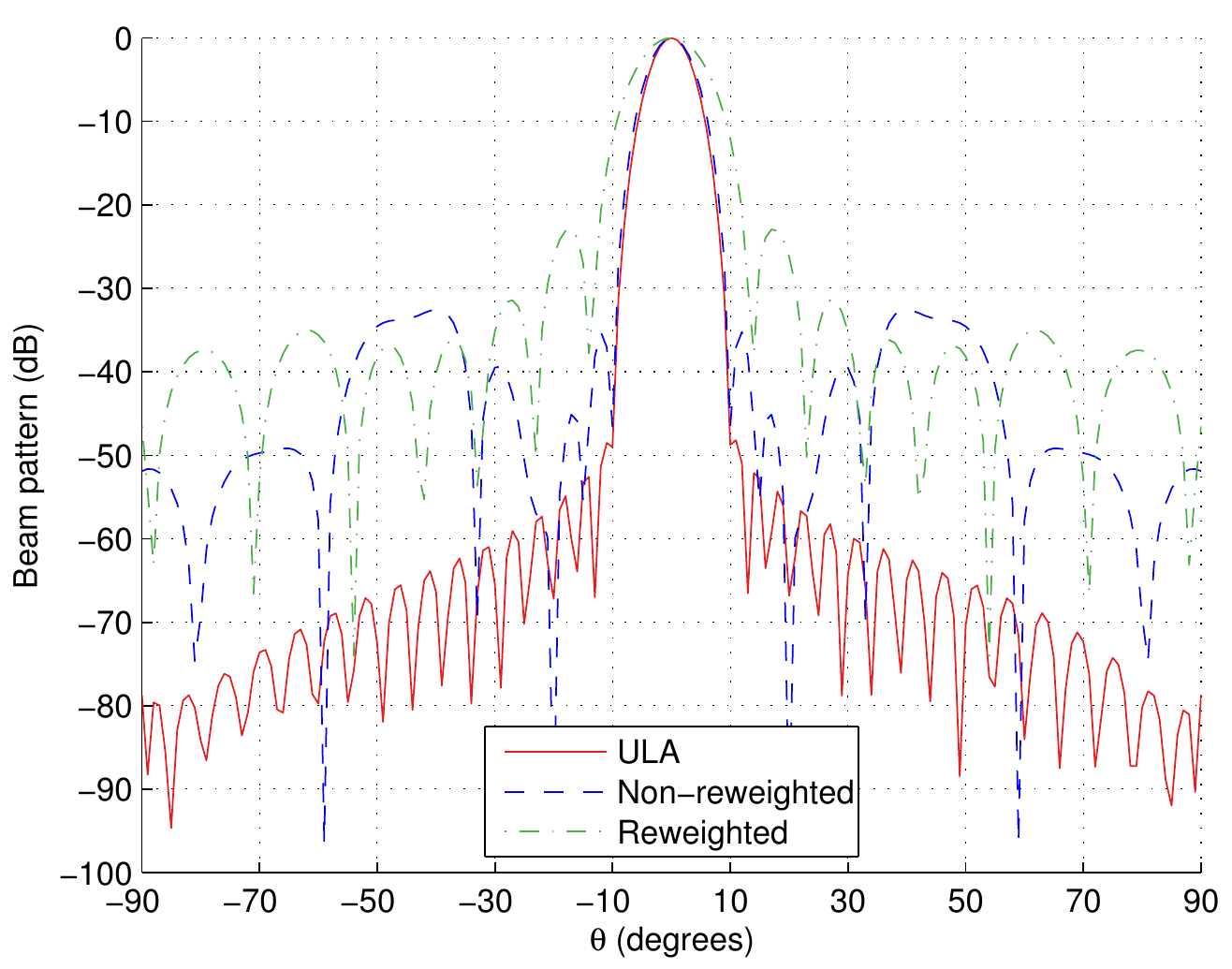}
   \caption{Responses for broadside design example, $M=301$.
    \label{fig:broad}}
\end{center}
\end{figure}

As an example, the tripole locations for each method when $M=301$ are shown in \mbox{Tables \ref{tb:broadnon} and \ref{tb:broadre},} respectively, with Figure \ref{fig:broad} showing the designed responses.  For completeness a ULA with adjacent tripole separation of $\lambda/2$ and aperture of $10\lambda$ is also shown for a comparison.  The weight coefficients for this comparison ULA can be found using the following:
\begin{eqnarray}\label{eq:ULA}\nonumber
    &\min&||\hat{\textbf{p}}_r-\hat{\textbf{S}}_{ULA}\hat{\textbf{w}}_{ULA}^{T}||_{2}\\ &\textrm{subject to}&
    \hat{\textbf{S}}_{ULA,\theta_{ML}}\hat{\textbf{w}}_{ULA}^{T}=1
\end{eqnarray}
where $\textbf{S}_{ULS}$ is given by
 \begin{equation}\label{eq:S_hat_ula}
    \hat{\textbf{S}}_{ULA}=\left(
                       \begin{array}{cc}
                         \textbf{R}(\textbf{s}_{ULA,x,1}) & \textbf{I}(\textbf{s}_{ULA,x,1})  \\
                         \textbf{R}(\textbf{s}_{ULA,y,1}) & \textbf{I}(\textbf{s}_{ULA,y,1})  \\
                         \textbf{R}(\textbf{s}_{ULA,z,1}) & \textbf{I}(\textbf{s}_{ULA,z,1})  \\
                         -\textbf{I}(\textbf{s}_{ULA,x,1}) & \textbf{R}(\textbf{s}_{ULA,x,1})  \\
                         -\textbf{I}(\textbf{s}_{ULA,y,1}) & \textbf{R}(\textbf{s}_{ULA,y,1})  \\
                         -\textbf{I}(\textbf{s}_{ULA,z,1}) & \textbf{R}(\textbf{s}_{ULA,z,1})  \\
                         \vdots & \vdots  \\
                         -\textbf{I}(\textbf{s}_{ULA,z,M}) & \textbf{R}(\textbf{s}_{ULA,z,M})  \\
                       \end{array}
                     \right)^{T},
\end{equation}
$\textbf{s}_{ULA,i,m}$ is the contribution of the $m^{th}$ dipole parallel to the $i$-axis, $\hat{\textbf{S}}_{ULA,\theta_{ML}}$ is $\hat{\textbf{S}}_{ULA}$ just for \mbox{the mainlobe,}
 \begin{eqnarray}\label{eq:w_ula}\nonumber
   \hat{\textbf{w}}_{ULA} &=& [R(w_{ULA,x,1}), R(w_{ULA,y,1}), R(w_{ULA,z,1}), -I(w_{ULA,x,1}), -I(w_{ULA,y,1}),\\ && -I(w_{ULA,z,1}), \ldots, -I(w_{ULA,z,M})]
 \end{eqnarray}
 and $w_{ULA,i,m}$ is the weight coefficient for the $m^{th}$ dipole parallel to the $i$-axis.  We can view this as finding the weight coefficients $\hat{\textbf{w}}_{ULA}$ that minimise the difference between the reference and designed responses, subject to keeping a unity response for the mainlobe direction.  In practice, this has the effect of ensuring a distortionless response at the mainlobe while minimising the sidelobes.

\begin{table}[H]
\caption{\rm Locations for the non-reweighted broadside design example, $M=301$.} \centering
\begin{tabular}{ccccccccccccccc}
\textbf{n}  & \textbf{1} & \textbf{2} & \textbf{3}  & \textbf{4} & \textbf{5} & \textbf{6} & \textbf{7} & \textbf{8} & \textbf{9} & \textbf{10} & \textbf{11} & \textbf{12} & \textbf{13} & \textbf{14} \\
\hline
$d_{n}/\lambda$  & 0 & 0.80 & 1.62  & 2.42 & 3.25  & 4.02 & 4.72 & 5.28 &5.98 & 6.75&7.58 &8.38 &9.20 & 10   \\

\end{tabular}
\label{tb:broadnon}
\end{table}

\begin{table}[H]
\caption{\rm Locations for the reweighted broadside design example, $M=301$.} \centering
\begin{tabular}{ccccccccc}
\textbf{n}  & \textbf{1} & \textbf{2} & \textbf{3}  & \textbf{4} & \textbf{5} & \textbf{6} & \textbf{7} & \textbf{8}  \\
\hline
$d_{n}/\lambda$  & 2.43 & 3.23 & 4.03  & 4.70 & 5.30  & 5.97 & 6.77 & 7.57    \\

\end{tabular}
\label{tb:broadre}
\end{table}

Note, Figure \ref{fig:broad} shows that the responses for the arrays designed using the two proposed methods are slightly less desirable than the comparison ULA.  This is to be expected, as significantly less tripoles are being used (21 for the ULA compared to 14 or 8 for the proposed methods).  However, they still give acceptable approximations of the reference response.  Here, the important thing is the reduction in the number of tripoles as was the aim of the proposed methods.
\subsection{Off-Broadside Design Examples}

We will now present off-broadside design examples for both of the proposed methods.  Here, we will use the parameters from the broadside design example but with the value of $M=301$ fixed.  The performance will be considered over a range of desired mainlobe locations ($\theta_{ML}$ varies for a fixed $\phi_{ML}=90^{\circ}$) and, in each case, there will be a transition region of $10^{\circ}$ on either side of the mainlobe with the remaining regions being the sidelobes.

Table \ref{tb:offbroadcompare} summarises the performance of the proposed methods over the range of mainlobes being considered.  We can see that an acceptable approximation of the reference response is achieved in all cases (\textit{i.e.}, mainlobe is within $1^{\circ}$ of what is designed and sufficient sidelobe attenuation is achieved).  In addition, for each example, using the reweighted method has reduced the number of tripoles required, but at the expense of an increased computation time ($321\%$ for $\theta=10^{\circ}$, $295\%$ for $\theta=30^{\circ}$, $582\%$ for $\theta=50^{\circ}$ and $544\%$  for $\theta=70^{\circ}$).  In the case of the $\theta_{ML}=50^{\circ}$ and $\theta_{ML}=70^{\circ}$ examples, this reduction in tripoles has even made the difference between whether there are any savings in number of tripoles as compared to an equivalent length ULA or not.  However, for the non-reweighted method, we can also get a result with improved sparsity by increasing the value of $\alpha$, although this will mean that the designed response will be a worse approximation of the reference response as a result.  For example, see Table \ref{tb:increased}, where a value of $\alpha=0.95$ is shown to give a reduction in the number tripoles required when using the non-reweighted method.

\begin{table}[H]
\caption{\rm Off-broadside performance comparison, for $\alpha=0.5$.} \centering
\small
\begin{tabular}{cccccccc}
  &  & \textbf{Number of} &  & &\textbf{Computation}   & &\\
\boldmath$\theta_{ML}$&\boldmath\textbf{Aperture}$/\lambda$ & \boldmath\textbf{Tripoles}& \boldmath$\overline{\Delta\textrm{$d$}}/\lambda$& \boldmath\textbf{Min(}$\Delta\textrm{$d$}$\textbf{)}& \textbf{Time (Seconds)}&\boldmath$\hat{\theta}_{ML}$ & \textbf{PSL (dB)}\\
\hline

 Non-reweighted $10^{\circ}$& 10 &  15& 0.71& 0.68&   22.17   & $10^{\circ}$&$-$35.18\\
\hline

Reweighted $10^{\circ}$  &  5.73& 9 & 0.72&0.70& 71.17  &$10^{\circ}$ &$-$21.27\\
\hline

 Non-reweighted $30^{\circ}$& 10 & 18 &0.59 & 0.55&   23.40   &$30^{\circ} $&$-$30.92\\
\hline

Reweighted $30^{\circ}$  & 6.53 & 12 & 0.59&0.57&  69.05 &$29^{\circ}$ &$-$23.27\\
\hline

  Non-reweighted $50^{\circ}$& 10 & 23 & 0.45& 0.35& 24.88     & $50^{\circ}$&$-$24.69\\
\hline

Reweighted $50^{\circ}$  & 8.20 & 6 &1.64 &1.57&  144.89 & $49^{\circ}$&$-$18.12\\
\hline

 Non-reweighted $70^{\circ}$&10  & 24 &0.43 &0.17 & 21.18     &$69^{\circ}$ &$-$19.61\\
\hline

Reweighted $70^{\circ}$  & 10 & 8 &1.43 &0.17& 115.22  &$70^{\circ}$ &$-$17.51\\

\end{tabular}
\label{tb:offbroadcompare}
\end{table}

\begin{table}[H]
\caption{\rm Off-broadside performance comparison, for $\alpha=0.95$.} \centering
\small
\begin{tabular}{cccccccc}
  &  & \textbf{Number of} &  & &\textbf{Computation}   & &\\
\boldmath$\theta_{ML}$&\boldmath\textbf{Aperture}$/\lambda$ & \textbf{Tripoles}& \boldmath$\overline{\Delta\textrm{$d$}}/\lambda$& \boldmath\textbf{Min(}$\Delta\textrm{$d$}$\textbf{)}& \textbf{Time (Seconds)}&\boldmath$\hat{\theta}_{ML}$ & \textbf{PSL (dB)}\\
\hline

 Non-reweighted $50^{\circ}$& 10 & 18 & 0.59 & 0.48&    25.25  & $50^{\circ}$&$-$15.23\\
\hline

Non-reweighted $70^{\circ}$  & 10& 20 & 0.53&0.33& 25.44  &$70^{\circ}$ &$-$7.59\\
\hline
\end{tabular}
\label{tb:increased}
\end{table}

\begin{figure}[H]
\begin{center}
   \includegraphics[angle=0,width=.6\textwidth]{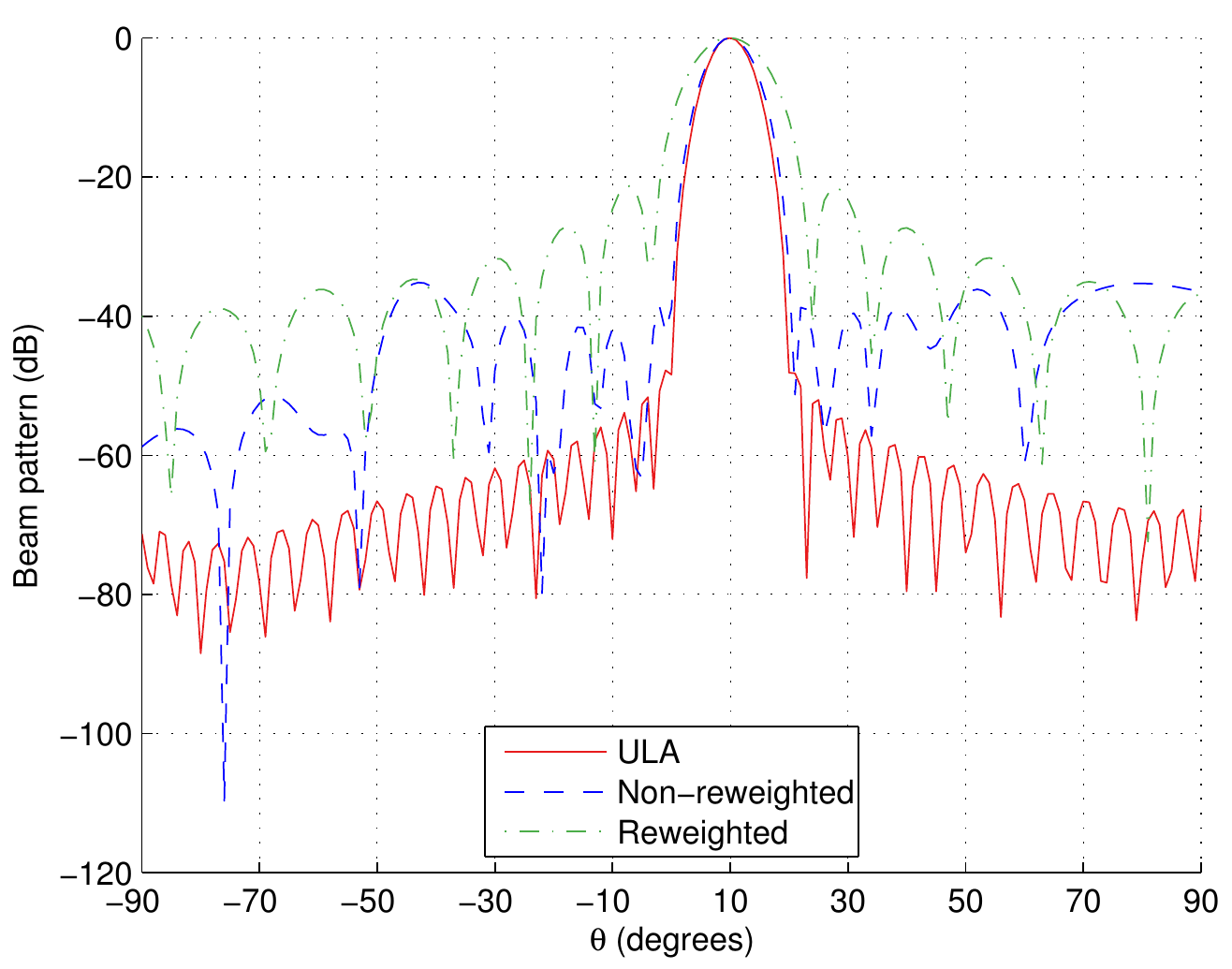}
   \caption{Responses for off-broadside design example, $\theta_{ML}=10^{\circ}$.
    \label{fig:offbroad}}
\end{center}
\end{figure}

Here, we show the responses for $\theta_{ML}=10^{\circ}$ as an example in Figure \ref{fig:offbroad}.  Again, the response for a ULA of aperture $10\lambda$ and adjacent tripole separation of $\lambda/2$ is shown for a comparison. As for the broadside example with $M=301$, we can see that the designed responses are again slightly less desirable (wider mainlobe and increased PSL).  However, this is an acceptable cost for achieving the desired reduction in the number of tripoles used.  For completeness, the resulting tripole locations are shown in Tables \ref{tb:offbroadnon} and \ref{tb:offbroadre}, respectively.

\begin{table}[H]
\caption{\rm Locations for the non-reweighted off-broadside design example, $\theta_{M}=10^{\circ}$.} \centering
\small
\begin{tabular}{cccccccccccccccc}
\textbf{n}  & \textbf{1} & \textbf{2} & \textbf{3}  & \textbf{4} & \textbf{5} & \textbf{6} & \textbf{7} & \textbf{8} & \textbf{9} & \textbf{10} & \textbf{11} & \textbf{12} & \textbf{13} & \textbf{14} & \textbf{15} \\
\hline
$d_{n}/\lambda$  & 0 & 0.68 & 1.40  &  2.12&2.85   &  3.55& 4.28 & 5 & 5.72& 6.45& 7.15&7.88 & 8.60& 9.32 & 10  \\

\end{tabular}
\label{tb:offbroadnon}
\end{table}

\begin{table}[H]
\caption{\rm Locations for the reweighted off-broadside design example, $\theta_{M}=10^{\circ}$.} \centering
\begin{tabular}{cccccccccc}
\textbf{n}  & \textbf{1} & \textbf{2} & \textbf{3}  & \textbf{4} & \textbf{5} & \textbf{6} & \textbf{7} & \textbf{8}  &\textbf{9}\\
\hline
$d_{n}/\lambda$  & 2.13 &  2.83&   3.57& 4.27 & 5  & 5.73 & 6.43 & 7.17& 7.87   \\

\end{tabular}
\label{tb:offbroadre}
\end{table}

%%%%%%%%%%%%%%%%%%%%%%%%%%%%%%%%%%%%%%%%%%

\section{Conclusions}\label{sec:con}

This paper has addressed the problem of designing sparse linear tripole arrays for the first time.  Unlike for isotropic array elements, each antenna now has three complex valued weight coefficients (one for each co-located dipole) which have to be simultaneously minimised in order to guarantee a truly sparse solution.  Therefore, we have proposed two CS-based solutions to this problem: (i) a modified $l_{1}$ norm minimisation; (ii) a series of iteratively solved reweighted minimisations, which give an improved level of sparsity.  Both broadside and off-broadside design examples have been provided to verify the effectiveness of the proposed methods.  In both cases, the results show that a good approximation of the reference pattern can be achieved using fewer tripoles than a ULA of equivalent length.  Note, in this paper, we have focused on the case of a single polarisation of interest, as we were more interested in showing it was possible to design sparse tripole arrays using CS rather than dealing with multiple polarisations.  In order to handle multiple polarisations, it may be necessary to extend the proposed methods to designing planar arrays, which is seen as an area for future research.

%%%%%%%%%%%%%%%%%%%%%%%%%%%%%%%%%%%%%%%%%%

\section*{Acknowledgments} \noindent We appreciate the support of the UK Engineering and Physical Sciences Research Council (EPSRC) via the project Bayesian Tracking and Reasoning over Time (BTaRoT) grant EP/K021516/1.

%%%%%%%%%%%%%%%%%%%%%%%%%%%%%%%%%%%%%%%%%%
\section*{Author Contributions} \noindent
Matthew Hawes developed the proposed design methods and wrote the manuscript. Wei Liu formulated the original problem being considered.  Wei Liu and Lyudmila Mihaylova were involved in the theoretical development of the solutions, in editing and structuring the manuscript.  All authors were involved in discussing potential solutions, the results and determining the best test scenarios to be used.

%%%%%%%%%%%%%%%%%%%%%%%%%%%%%%%%%%%%%%%%%%
\section*{Conflicts of Interest} \noindent
The authors declare no conflict of interest.

%=================================================================
% References: Variant A
%=================================================================
% Back Matter (References and Notes)
%----------------------------------------------------------
% Style and layout of the references
\bibliographystyle{IEEEtran}
\bibliography{mybib}
%%MDPI internal note: new layout%% redefinition removed

\end{document}